# Hierarchical Design Based Intrusion Detection System For Wireless Ad hoc Sensor Network

Mohammad Saiful Islam Mamun
Department of Computer Science, Stamford University Bangladesh, 51, Siddeshwari, Dhaka. E-mail : msimamun@kth.se

A.F.M. Sultanul Kabir
Department of Computer Science and Engineering, American International University Bangladesh, Dhaka. afmk@kth.se

*ABSTRACT*

*In recent years, wireless ad hoc sensor network becomes popular both in civil and military jobs. However, security is one of the significant challenges for sensor network because of their deployment in open and unprotected environment. As cryptographic mechanism is not enough to protect sensor network from external attacks, intrusion detection system needs to be introduced. Though intrusion prevention mechanism is one of the major and efficient methods against attacks, but there might be some attacks for which prevention method is not known. Besides preventing the system from some known attacks, intrusion detection system gather necessary information related to attack technique and help in the development of intrusion prevention system. In addition to reviewing the present attacks available in wireless sensor network this paper examines the current efforts to intrusion detection system against wireless sensor network. In this paper we propose a hierarchical architectural design based intrusion detection system that fits the current demands and restrictions of wireless ad hoc sensor network. In this proposed intrusion detection system architecture we followed clustering mechanism to build a four level hierarchical network which enhances network scalability to large geographical area and use both anomaly and misuse detection techniques for intrusion detection. We introduce policy based detection mechanism as well as intrusion response together with GSM cell concept for intrusion detection architecture.*

## 1. INTRODUCTION

There has been a lot of research done on preventing or defending WSN from attackers and intruders, but very limited work has been done for detection purpose. It will be difficult for the network administrator to be aware of intrusions. There are some Intrusion Detection Systems that are proposed or designed for Wireless Ad hoc network. Most of them work on distributed environment; which means they work on individual nodes independently and try to detect intrusion by studying abnormalities in their neighbors' behavior. Thus, they require the nodes to consume more of their processing power, battery backup, and storage space which turn IDS to be more expensive, or become unfeasible for most of the applications. Some of the IDS use mobile agents in distributed environment [8]. Mobile Agent supports sensor mobility, intelligent routing of intrusion data throughout the network, eliminates network dependency of specific nodes. But this mechanism still is not popular for IDS due to mobile agents' architectural inherited security vulnerability and heavy weight. Some of the IDSs are attack-specific which make them concentrated to one type of attack [1] . Some of them use centralized framework which make IDS capable exploiting a personal computer's high processing power, huge storage capabilities and unlimited battery back up [21]. Most of the IDS are targeted to routing layer only [7] [21], but it can be enhanced to detect different types of attacks at other networking layers as well. Most of the architectures are based on anomaly detection [18] [2] which examine the statistical analysis of activities of nodes for detection. Most of the IDS techniques utilize system log files, network traffic or packets in the network to gather information for Intrusion detection. Some detects only intrusion and some do more like acquiring more information e.g. type of attacks, locations of the intruder etc. Though a handsome number of IDS mechanisms are proposed in Wireless ad hoc network but very few of them can be applicable for Wireless Sensor network because of their resource constrains. Self-Organized Criticality & Stochastic Learning based IDS [2], IDS for clustering based sensor Networks [3], A non-cooperative game approach [4], Decentralized IDS [5] are distinguished among them.

## 2. EXISTING CHALLENGES

Existing intrusion detection systems are not adequate to protect WSN from Inside and Outside attackers. None of them are complete. E.g. most of the approaches offer clustering techniques without mentioning how they will be formed and how will they behave with rest of the system. Most of the existing IDSs deal with wired architecture except their wireless counterpart. The architecture of WSN is even more sophisticated than ad hoc wireless architecture. So, an IDS is needed with capability of detecting inside and outside, known and unknown attacks with low false alarm rate. Existing IDS architecture that are specifically designed for sensor networks are suffering from lack of resources e.g. high processing power, huge storage capabilities, unlimited battery backup etc.

## 3. WIRELESS SENSOR NETWORK - AN OVERVIEW

According to NIST (National Institute of Standards and Technology) "*a wireless ad hoc sensor network consists of a number of sensors spread across a geographical area*" [8]. The term *sensor network* refers to a system which is a combination of sensors and actuators with some general purpose computing elements. A sensor network can have hundreds or even thousands of sensors; mobile or fixed locations; deployed to control or monitor [7].

A wireless sensor network comprises of sensor nodes to sense data from their ambience, and passes it on to a centralized controlling and data collecting identity called *base station*. Typically, base stations are powerful devices with a large storage capacity to store incoming data. They generally provide gateway functionality to another network, or an access point for human interface [21]. A base station may have an unlimited power supply and high bandwidth links for communicating with other base stations. In contrast, wireless sensors nodes are constrained to use low power, low bandwidth, and short range links.

## 4. SECURITY THREATS AND ISSUES

Various security issues and threats that are considered for wireless ad hoc network can be applied for WSN. This is recited in some previous researches. But the security mechanism used for wireless ad hoc networks cannot be deployed directly for WSNs because of their architectural inequality. First, in ad hoc network, every node is usually held and managed by a human user. Whereas in sensor network, all the nodes are independent and communication is controlled by base station. Second, Computing resources and batteries are more constrained in sensor nodes than in ad hoc nodes. Third, the purpose of sensor networks is very specific e.g. measuring the physical information (such as temperature, sound etc.). Fourth, node density in sensor networks is higher than in ad hoc networks [10]. Architectural aspect of WSN makes the security mechanism more prosperous as the base station could be used intelligently.

According to the basic need of security attacks in WSN can be categorized:

- DoS, DDoS attacks which affect network *availability*
- Eavesdropping, sniffing which can threaten *confidentiality*
- Man-in-the-middle attacks which can affect packet *integrity*
- Signal jamming which affects *communication*

There are many research work has been done in the area of significant security problems. Here summery of existing well-known threats are discussed.

Table 1: **Threats and Attacks in WSN**

| **Attacks** | **Brief Description** |
|---|---|
| Attack on Information in transit | Information that is to be sent can be modified, altered, replayed, spoofed, or vanished by attacker. |
| Hello flood | Attacker with high radio range sends more Hello packet to announce themselves to large number of nodes in the large network persuading themselves as neighbor. |
| Sybil attack | Fake multiple identities to attack on data integrity and accessibility. |
| Wormhole attack | Transmit information between two WSN nodes in secret. |
| Network partition attack | Threats to accessibility though there is a path between the nodes. |
| Black Hole Attack | The attacker absorbs all the messages. |
| Sink Hole Attack | Similar to black hole. Exception: the attacker advertises wrong routing information |
| Selective Forwarding | The attacker forwards messages on the basis of some Pre-selected criterion |
| Simple Broadcast Flooding | The attacker floods the network with broadcast Messages. |
| Simple Target Flooding | The attacker tries to flood through some specific nodes. |
| False Identity Broadcast Flooding | Similar to simple broadcast flooding, except the attacker deceives with wrong source ID. |
| False Identity Target Flooding | Similar to simple target flooding, except the attacker deceives with wrong source ID. |
| Misdirection Attack | The attacker misdirects the incoming packets to a distant node. |

## 5. IDS ARCHITECTURE

According to the Network Security Bible – *"Intrusion detection and response is the task of monitoring systems for evidence of intrusions or inappropriate usage and responding to this evidence"*[22]. The basic idea of IDS is to observe user as well as program activities inside the system via auditing mechanism.

Depending on the data collection mechanism IDS can be classified into two categories: *Host based IDS* monitors log files (applications, Operating system etc.) and then compare with logs of present signature of known attacks from internal database. *Network based IDS* works in different way. It monitors packets within communication and inspects suspicious packet information.

Depending on how attacks are detected, IDS architecture can be categorized into three types: *Signature based IDS* which monitors an occurrence of signatures or behaviors which is matched with known attacks to detect an intrusion. This technique may exhibit low false positive rate, but not good to detect previously unknown attacks. *Anomaly based IDS* defines a profile of normal behavior and classifies any deviation of that profile as an intrusion. The normal profile of system behavior is updated as the system learns the behavior. This type of system can detect unknown attacks but it exhibits high false positive. In [11] another type of Intrusion detection has been introduced. *Specification based IDS* defines a protocol or a program's correct operations. Intrusion is indicated according to those constraints. This type of IDS may detect unknown attacks, while showing low false positive rate.

In [11] wireless ad hoc network architecture is defined into three basic categories which can be adjusted to IDS in WSN architecture.

*Stand alone*
Each node acts as an independent IDS and detects attacks for itself only without sharing any information with another IDS node of the system, even does not cooperate with other systems. So, all intrusion detection decisions are based on information available to the individual node. Its effect is too limited. This architecture is best suited in an environment where all the nodes are capable of running an IDS [11].

*Distributed and cooperative*
Though each node runs its own IDS, finally they collaborate to form a global IDS. This architecture is more suitable for flat wireless sensor networks, where a global IDS is initiated due to the occurrence of inconclusive intrusions detected by individual node.

*Hierarchical*
This architecture has been proposed for multilayered wireless network. Here network is divided into cluster with cluster-heads. Cluster-head acts like a small base station for the nodes within the cluster. It also aggregates information from the member nodes about malicious activities. Cluster-head detects attacks as member-nodes could potentially reroute, modify or drop packet in transmission. At the same time all cluster-heads can cooperate with central base station to form a global IDS.

To build an effective IDS model, several considerations take place.
**First** of all Detection Tasks: How will they be separated? Local agent or Global agent. Whether Local or global agent, an IDS needs to consider how these agents would analyze the threats. And what would be right sources of information?
*Local Agent* detects vulnerability of node's internal Information. It supposed to be active 100% of the time to ensure maximum security. Here Physical/Logical Integrity,
Measurement Integrity, Protocol Integrity, Neighborhood are analyzed from nodes' status.
*Global Agent:* To detect anomaly from external information of a node to achieve 100% coverage of a sensor network. Here main challenges are balancing tasks and network coverage. In case of hierarchical network, cluster head (CH) controls its section of the network. CH is the part of global network. In case of flat network Spontaneous Watchdogs concept is applied. Here premise is "For every packet circulating in the network, there are a set of nodes that are able to receive both that packet and the relayed packet by the next hop."

**Second** consideration is Sharing Information between agents. Information between agents can be transmitted through cryptography, voting mechanism or trust depending on the network's resource constraint.

**Third** consideration is how to Notify Users. Generally users are behind Base stations. So, different algorithms can be used to notify base station. E.g. uTesla use secure broadcast algorithm.

There are different techniques for IDS in Wireless Sensor Network (WSN). Here we represent some existing IDS models for WSN.

Table 2: Comparative study on existing IDS

| Name of the Intrusion Detection System | Data Collection Mechanism | Detection technique | Handled attacks | Network Architecture |
|---|---|---|---|---|
| Hybrid IDS for Wireless Sensor Network [6] | Network based | Anomaly based | Selective forwarding, sink hole, Hello flood and wormhole attacks | Hierarchical |
| Decentralized IDS in WSN[5] | Network based | Anomaly based | Repetition, Message Delay, Blackhole, Wormhole, Data alteration, Jamming, Message negligence, and Selective Forwarding | Distributed |
| Intrusion Detection in Routing attacks in Sensor Network [1] | Host based | Anomaly based | DoS, active sinkhole attacks, and passive sinkhole. | Distributed |
| Sensor Network Automated Intrusion Detection System (SNAIDS)[9] | Host based | Signature based | duplicate nodes, flooding, Black hole, Sink hole attack, selective forwarding, misdirection. | Distributed |
| Self-Organized critically & stochastic learning based IDS for WSN[2] | Host based | Anomaly based | There is no guideline in this IDS model of which attacks it can resist and which cannot. | Distributed |

## 6. OUR MODEL

In this paper we propose a new model for IDS which concentrates on saving the power of sensor nodes by distributing the responsibility of intrusion detection to three layer nodes with the help of policy based network management system. The model uses a hierarchical overlay design(HOD). We divided each area of sensor nodes into hexagonal region (like GSM cells). Sensor nodes in each of the hexagonal area are monitored by a cluster node. Each cluster node is then monitored by a regional node. In turn, Regional nodes will be controlled and monitored by the Base station.

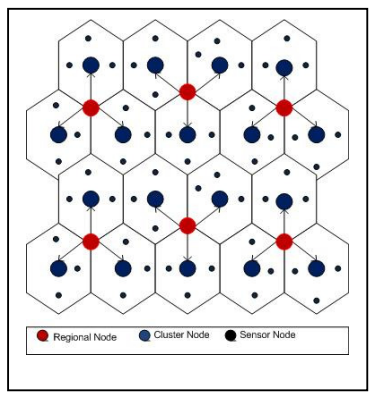
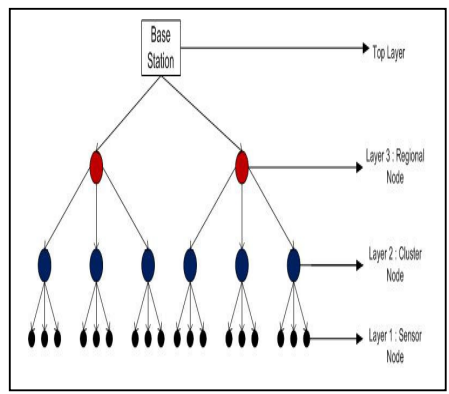

Figure 1: Hierarchical Overlay Design

This HOD based IDS combines two approaches of intrusion detection mechanisms (Signature and anomaly) together to fight against existing threats. Signatures of well known attacks are propagated from the base station to the leaf level node for detection. Signature repository at each layer is updated as new forms of attacks are found in the system. As intermediate agents are activated with predefined rules of system behavior, anomaly detection can take part from the deviated behavior of predefined specification. Thus proposed IDS can identify known as well as unknown attacks.

## 6.1 Detection Entities

*Sensor Nodes* have two types of functionality: Sensing and Routing. Each of the sensor nodes will sense the environment and exchange data in between sensor nodes and cluster node. As sensor nodes have much resource constraints, in this model, there is no IDS module installed in the leaf level sensor nodes.

*Cluster Node* plays as a monitor node for the sensor nodes. One cluster node is assigned for each of the hexagonal area. It will receive the data from sensor nodes, analyze and aggregate the information and send it to regional node. It is more powerful than sensor nodes and has intrusion detection capability built into it.

*Regional Node* will monitor and receive the data from neighboring cluster heads and send the combined alarm to the upper layer base station. It is also a monitor node like the cluster nodes with all the IDS functionalities. It makes the sensor network more scalable. If thousands of sensor nodes are available at the leaf level then the whole area will be split into several regions.

*Base Station* is the topmost part of architecture empowered with human support. It will receive the information from Regional nodes and distribute the information to the users based on their demand.

## 6.2 Policy based IDS

Policy implies predefined action pattern that is repeated by an entity whenever certain conditions occur [13]. The architectural components of policy framework include a Policy Enforcement Point (PEP), Policy Decision Point (PDP), and a Policy repository. The policy rules stored in Policy repository are used by PDP to define rules or to show results. PDP translates or interprets the available data to a device-dependent format and configures the relevant PEPs. The PEP executes the logical entities that are decided by PDP [12]. These capabilities provide powerful functions to configure the network as well as to re-configure the system as necessary to response to network conditions with automation. In a large WSN where Hierarchical Network Management is followed can be realized by policy mechanism to achieve survivability, scalability and autonomy simultaneously. So in case of failure the system enables one component to take over the management role of another component. One of the major architectural advantages of hierarchical structure is any node can take over the functionality of another node dynamically to ensure survivability. A flexible agent structure ensures dynamic insertion of new management functionality.

*Hierarchical network* management integrates the advantage of two (Central and Distributed) management models [14] and uses intermediate nodes (Regional and Cluster) to distribute the detection tasks. Each intermediate manager has its own domain called Regional or Cluster agent which collects and processed information from its domain and passes the required information to the upper layer manager for further steps. All the intermediate nodes are also used to distribute command/data/message from the upper layer manager to nodes within its domain. It should be noted that there is no direct communication between the intermediate members. Except the leaf level sensor nodes all the nodes in the higher level are configured with higher energy and storage.

To achieve a policy-based management for IDS the proposed architecture features several components that evaluate policies: a Base Policy decision Point (BPDP), a number of Policy decision modules (PDMs) and Policy Enforcement Point(PEP).

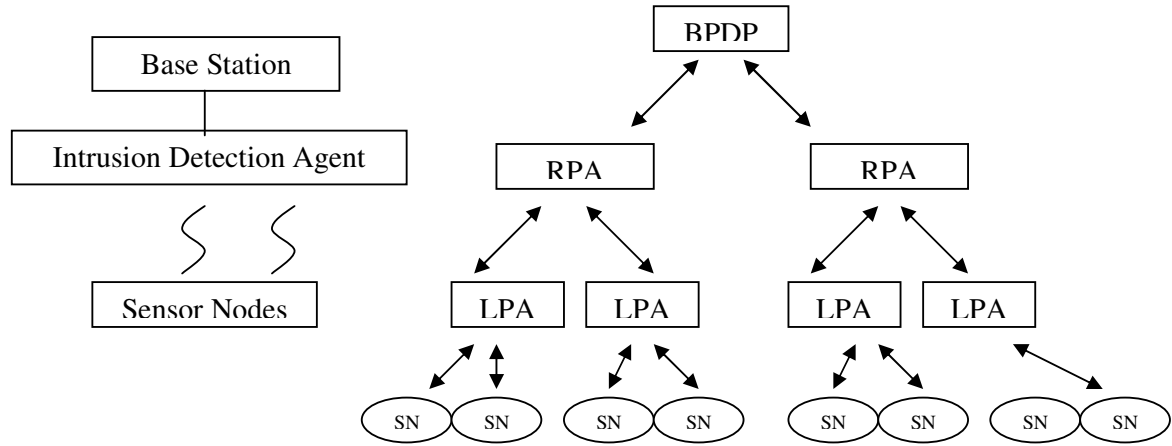

BPDP: Base Policy Decision Point

RPA: Regional Policy Agent

LPA: Local Policy Agent

SN: Sensor Node

Figure 2: Hierarchical Architecture of IDS Policy Management

*Base Policy Decision Point (BPDP)* is the controlling component of the architecture. It implements policies or intrusion rules generated by the Intrusion Detection Tool (IDT) from receiving events, evaluating anomaly conditions and applying new rules, algorithms, threshold values etc. IDT supports creation, deletion, modification, and examination of the agent's configurations and policies. It can add new entities e.g. new signature of intrusion, modify or delete existing entities in RPA and LPA.

*Policy Decision Modules (PDMs)* are components that implement sophisticated algorithms in relevant domains. LPAs and RPAs act as PDMs. LPA manages the sensor nodes which is more powerful than sensor nodes. LPAs perform local policy-controlled configuration, filtering, monitoring, and reporting which reduces management bandwidth and computational overhead from leaf level sensor nodes to improve network performance and intrusion detection efficiency. An RPA can manage multiple LPAs. At the peak BPDP manages and controls all the RPAs.

*Policy Enforcement Points* (PEP) are low level Sensor Nodes.

Policies are disseminated from the BPDP to RPA to LPA as they are propagated from PDP to LPA. Policy agents described above helps IDS by reacting to network status changes globally or locally. It helps the network to be reconfigured automatically to deal with fault and performance degradation according to intrusion response.

## 6.3 Structure of Intrusion Detection Agent ( IDA )

The hierarchical architecture of policy management for WSN is shown in the above figure. It comprises of several hierarchical layers containing Intrusion Detection Agent (IDA) at each layer. They are Base Policy Decision Point (BPDP), Regional Policy Agent( RPA), Local Policy Agent (LPA), Sensor Node(SN).

An IDA consists of the following components: Preprocessor, Signature Processor, Anomaly processor and Post processor. The functionalities are described as follows.

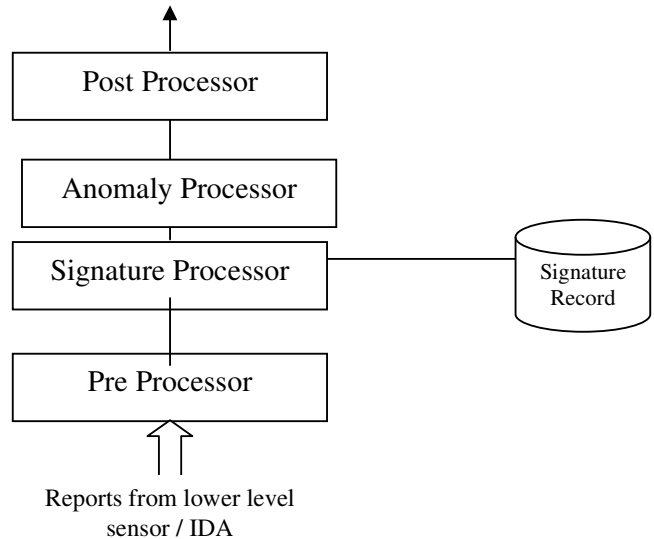

Figure 3: Intrusion Detection Agent Structure

*Pre-Processor* either collects the network traffic of the leaf level sensor when it acts as an LPA or it receives reports from lower layer IDA. Collected sensor traffic data is then abstracted to a set of variables called stimulus vector to make the network status understandable to the higher layer processor of the agent.

*Signature Processor* maintains a reference model or database called *Signature Record* of the typical known unauthorized malicious threats and high risk activities and compares the reports from the *preprocessor* against the known attack signatures. If match is not found then misuse intrusion is supposed to be detected and signature processor passes the relevant data to the next higher layer for further processing.

*Anomaly Processor* analyzes the vector from the *preprocessor* to detect anomaly in network traffic. Usually statistical method or artificial intelligence is used in order to detect this kind of attack. Profile of normal activity which is propagated from Base station is stored in the database. If the activities arrived from preprocessor deviates from the normal profile in a statistically significant way, or exceeds some particular threshold value attacks are noticed. Intrusion detection rules are basically policies which define the standard of access mechanism and uses of sensor nodes. Here database acts as a Policy Information Base(PIB) or policy repository.

*Post Processor* prepares and sends reports for the higher layer agent or base station. It can be used to display the agent status through a user interface.

## 6.5 Selection of IDS node

Activating every node as an IDS wastes energy. So minimization of number of nodes to run intrusion detection is necessary. In [15] three strategies are mentioned involving selection of Intrusion detection node.

*Core defense* selects IDS node around a centre point of a subset of network. It is assumed that no intruder break into the central station in any cluster. This type of model defends from the most inner part then retaliates to the outer area.

*Boundary defense* selects node along the boundary perimeter of the cluster. It provides defense on intruder attack from breaking into the cluster from outside area of the network.

*Distributed defense has* an agent node selection algorithm which follows voting algorithm from [16] in this model. Node selection procedure follows tree hierarchy.

Our model follows *Core Defense* strategy where cluster-head is the centre point to defend intruders. In core defense strategy ratio of alerted nodes and the total number of nodes in the network drops, this makes energy consumption very low which make it more economical in their use of energy as it shows least number of broadcast message in case of attack. It has strong defense in inner network. Here IDS needs to wait for intruder to reach the core area [16] which is one of the drawbacks of this strategy as nodes can be captured without notice.

## 6.6 IDS mechanism in sensor nodes

Intrusions could be detected at multiple layers in sensor nodes (physical, Link, network and application layer).
*In Physical layer* Jamming is the primary physical layer attack. Identifying jamming attack can be done by the Received Signal Strength Indicator (RSSI) [17] [18], the average time required to sense an idle channel (carrier sense time), and the packet delivery ratio (PDR). In case of wireless medium, received signal strength has relation with the distance between nodes. Node tampering and destruction are another physical layer attack that can be prevented by placing nodes in secured place. During the initialization process Cluster node's LPA will store the RSSI value for the communication between Cluster node to leaf level sensor nodes and sensor to sensor node. Later, at the time of monitoring, Anomaly processor in LPA will monitor whether the received value is unexpected. If yes, it will feedback RPA by generating appropriate alarm.

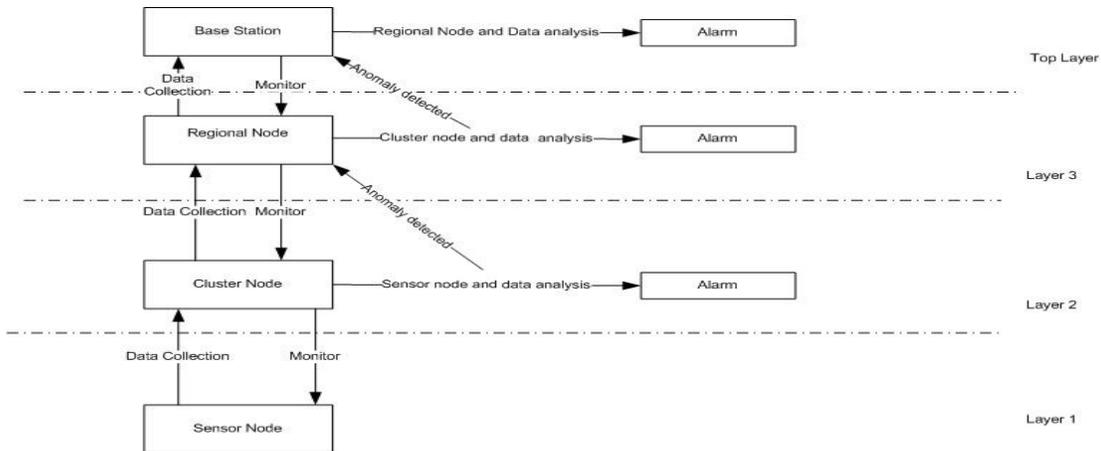

Figure 4: IDS mechanism

*Link Layer* attacks are collision, denial of sleep and packet replay etc. Here SMAC and Time Division Multiple Access (TDMA) can be used to detect the anomaly. *TDMA* [18] is digital transmission process where each cluster node will assign different time slots for different sensor nodes in its region. During this slot every sensor node has access to the radio frequency channel without interference. If any attacker send packet using source address of any node, e.g. A, but that slot is not allocated to A then LPA's Anomaly Processor can easily detect that intrusion. *S-MAC* [18] protocol is used to assign a wakeup and sleep time for the sensor nodes. As the sensor has limited power, S-MAC can be implemented for the energy conservation. If any packet is received from source e.g. A in its sleeping period then LPA can easily detect the inconsistency.

In *Network Layer* route tracing is used to detect whether the packet really comes from the best route. If packet comes to the destination via different path rather than the desired path then the Anomaly Processor can detect possible intrusion according to predefined rules.

*Application Layer* uses three level watchdogs. They are in base station, regional node, cluster node. Sensor nodes will be monitored by upper layer watchdog cluster node and cluster nodes will be monitored by regional node watchdog and finally the top level watchdog base station will monitor the regional nodes. So, if any one node is compromised by the attacker then higher layer watch dog can easily detect the attack and generate alarm.

## 7. INTRUSION RESPONSE

There are differences between intrusion detection and intrusion prevention. If a system has intrusion prevention, it is assumed that intrusion detection is built in. IDSs are designed to welcome intrusion to get into system; where as Intrusion Prevention System (IPS) actually attempts to prevent access to the system from the very beginning. IPS operates similar to IDS with one critical difference: "IPS can block the attack itself; while an IDS sits outside the line of traffic and observes, an IPS sits directly in line of network traffic. Any traffic the IPS identifies as malicious is prevented from entering the network [19]." So in case of IDS "Intrusion Response" should be the right title for recovery.

There are two different approaches for intrusion response: Hot response or Policy based response [20]. *Hot response* reacts by launching local action on the target machine to end process, or on the target network component to block traffic. E.g. kill any process, Reset connection etc. It does not prevent the occurrence of the attack in future. On the other hand *Policy based response* works on more general scope. It considers the threats reported in the alert, constraints and objectives of the information system of the network. It modifies or creates new rules in the policy repository to prevent an attack in the future. In our proposed IDS, Base station's Policy decision point and other policy decision modules take part in the response mechanism together. BPDP and PDM take part in response mechanism. Intrusion can be detected either in Cluster node or Regional node. Finally base stations can be involved anytime if network administrator wants to do so or to update signature database or policy stored in intermediate agent. Intrusions are detected automatically according to the policy implemented by BPDP. Re-action is also automatic but administrator may re-design the architecture according requirements.

In [21] a novel intrusion detection and response system is implemented. We have applied their idea in our response mechanism with some modification. Our IDS system considers each sensor nodes into one of five classes: *Fresh*, *Member*, *Unstable*, *Suspect* or *Malicious*. We have Local Policy Agent, Regional Policy Agent and finally Base Policy Decision Point to take decision about the sensor node's class placement. Routeguard mechanism use *Pathrating* algorithm to keep any node within these five classes [21]. In our model, we have policy or rules defined in Base station's BPDP to select any node to be within these five classes as shown in figure 4. When a new node is arrived, it will be classified as *Fresh*. For a pre-selected period of time this new node will be in *Fresh* state. By this time LPA will check whether this node is misbehaving or not. In this period the node is permitted to forward or receive packets from another sensor node, but not its own generated packet. After particular time its classification will be changed to Member automatically if no misbehave is detected. Otherwise the node's classification will be changed to *Suspect* state. In *Member* state nodes are allowed to create, send, receive or forward packets. In this time Member nodes are monitored by Watchdog at LPA in Cluster node. If the node misbehaves its state will be changed to *Unstable* for short span of time. During *Unstable* state nodes are permitted to send and receive packets except their own packets. In this state the node will be kept under close observation of LPA. If it behaves well then it will be transferred to *Member* state. A node in *Unstable* state will be converted to *Suspect* state in two cases: Either the node was in *Unstable* state and interchanged its state within *Member* and *Unstable* state for a particular amount of times (threshold value defined in LPA) within a predefined period or the node was misbehaving for long time (threshold value). LPA's Post processor sends "Danger alert" to RPA whenever *Suspect* node is encountered. The suspected node is completely isolated from the network. It is not allowed to send, receive, or forward packets and temporarily banned for short time. Any packets received from suspected node are simply discarded. After a certain period of time the node is reconnected and is monitored closely for extensive period of time by Intrusion Detection Agent in all three layers. If watchdogs report well then node status will be changed to *Unstable*. However if it continues misbehaving then it will be labeled as *Malicious*. After declaring any node malicious that node permanently banned from this network. To ensure that this malicious node will never try to reconnect, its MAC address or any unique ID will be added to *Signature Record Database* of LPA.

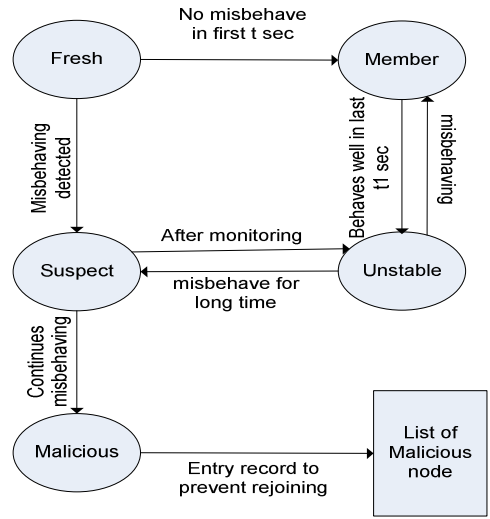

Figure 5 : Operation of Intrusion Response

Survivability is one of the major factors that are predicted from every system. We consider base stations to be failure free. But the Regional nodes or cluster nodes may be unreachable due to failure or battery exhaustion. So, in case of failures or any physical damage of Regional nodes or Cluster nodes, control of that node should be taken over by another stable node. So in our proposed architecture if any Regional node fails, then its control is shifted to the neighbor Regional node dynamically.

So, control of the Cluster nodes and sensor nodes belonging to that Regional node will be shifted automatically to the neighbor node. In the same way if any cluster node fails then control of that cluster node will be transferred to the neighbor Cluster node.

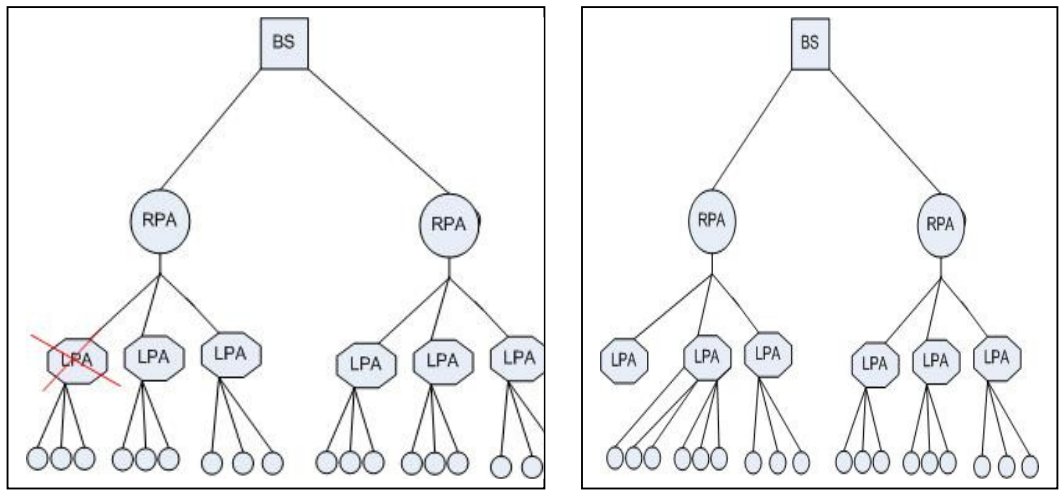

Figure 6: Cluster nodes failure

So in the proposed architecture if any LPA is unreachable due to failure or battery exhaustion of cluster nodes, neighbor LPA will take the charges of leaf level sensor nodes which was in the area of fault cluster node. In the same way due to Regional nodes failure neighbor Regional node's RPA will take over the functionality of all the cluster node's LPA and sensor nodes belonged to the faulty Regional node dynamically.

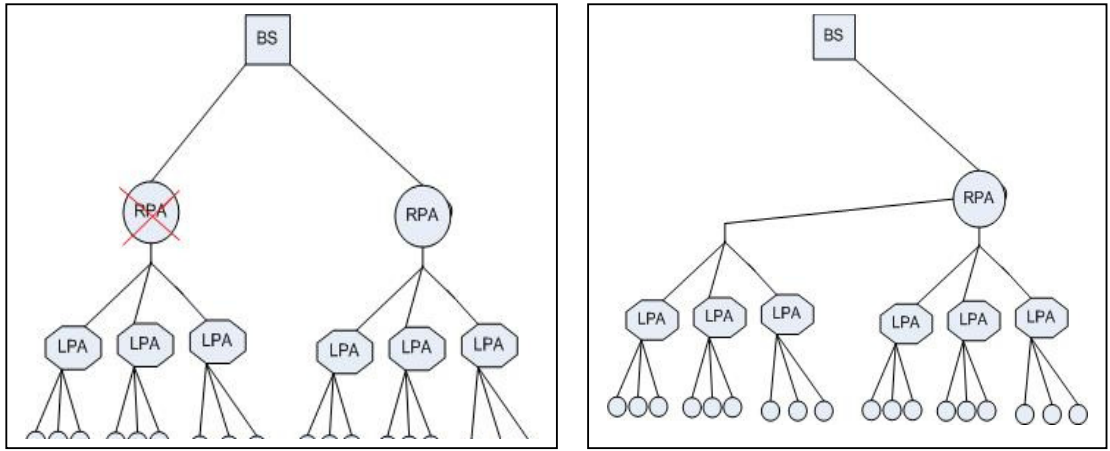

Figure 7 : Regional node's failure

As we mentioned before Cluster nodes or regional nodes havenumberdirect communications between them. So how will Cluster node or Regional node determine about the failure of its neighbor? Actually in the proposed architecture Base station has direct or indirect connections with all its leaf nodes. Base station has direct connection with Regional node. So if any Regional node fails Base station can identify the problem and select one of its neighbor nodes dynamically according to some predefined rule in BPDP. Then BPDP needs to supply the policy, rules, or signatures of failed node to the selected new neighbor Regional node. In the same way if any cluster node fails then neighbor cluster node will not be informed about its failure. So in this case Regional node will take necessary action of selecting suitable neighbor cluster node. Here policy, rules or signatures of the failed cluster node will be supplied by the BPDP through relevant RPA. So RPA has the only responsibility to select appropriate neighbor LPA of unreachable LPA. The rest of the work belongs to BPDP of Base station. As Base station is much more powerful node with large storage; all the signatures, anomaly detection rules or policies are stored primarily as backup in Base station. This back up system increases reliability of the whole network system.

## 8. CONCLUSION

WSN are prone to intrusions and security threats. In this paper, we propose a novel architecture of IDS for ad hoc sensor network based on hierarchical overlay design. We propose a response mechanism also according to proposed architecture. Our design of IDS improves on other related designs in the way it distributes the total task of detecting intrusion. Our model decouples the total work of intrusion detection into a four level hierarchy which results in a highly energy saving structure. Each monitor needs to monitor only a few nodes within its range and thus needs not spend much power for it. Due to the hierarchical model, the detection system works in a very structured way and can detect any intrusion effectively. As a whole, every area is commanded by one cluster head so the detection is really fast and the alarm is rippled to the base station via the region head enabling it to take proper action. In this paper we consider cluster nodes or Regional nodes to be more powerful than ordinary sensor nodes. Though it will increase the total cost of network set up, but to enhance reliability, efficiency and effectiveness of IDS for a large geographical area where thousands of sensor nodes take place, the cost is tolerable.

Policy based mechanism is a powerful approach to automating network management. The management system for intrusion detection and response system described in this paper shows that a well structured reduction in management traffic can be achievable by policy management. This policy-based architecture upgrades adaptability and re-configurability of network management system which has a good practical research value for large geographically distributed network environment.

The IDS in wireless sensor network is an important topic for the research area. Still there are no proper IDS in WSN field. Many previous proposed systems were based on three layer architecture. But we introduced a four layer overlay hierarchical design to improve the detection process and we brought GSM cell concept. We also introduced hierarchical watch dog concept. Top layer base station, cluster node and regional node are three hierarchical watchdogs. Our report proposes IDS in multiple layers to make our system architecture robust.

## 9. FUTURE WORK

This paper provides a first-cut solution to four layer hierarchical policy based intrusion detection system for WSN. So there is much room for further research in this area. Proposed IDS system is highly extensible, in that as new attack or attack pattern are identified, new detection algorithm can be incorporated to policy. Possible venues for future works include:

- Present model can be extended by exploring the secure communication between base station, Regional node and cluster node.
- The setting of management functions of manager station more precisely.
- Election procedure to select cluster and regional node: Instead of choosing the cluster node and regional node manually, there will be an election process that will automatically detect the cluster node and regional node.
- Implementation of Risk Assessment System in the manager stations to improve the reaction capability of intrusion detection system.
- In this paper we actually focus on the general idea of architectural design for IDS and how a policy management system can be aggregated to the system. But an extensive work needs to be done to define Detection and Response policy as well.
- Overall, more comprehensive research is needed to measure the current efficiency of IDS, in terms of resources and policy, so that improvements of its future version(s) are possible.
- Further study is required to determine IDS scalability. To the best of knowledge, its scalability highly correlates with the scalability of the WSN application and the policy management in use.
- Building our own Simulator: As all the previous research were based on three layer architecture, so we are planning to create our own simulator that will simulate our four layer design.

International Journal of Network Security & Its Applications (IJNSA), Vol.2, No.3, July 2010[2] S. Doumit and D.P. Agrawal,"**Self-organized criticality & stochastic learning based intrusion detection system for wireless sensor network**", MILCOM 2003 - IEEE Military Communications Conference, vol. 22, no. 1, pp. 609-614, 2003

[3]. C.-C. Su, K.-M. Chang, Y.-H. Kuo, and M.- F. Horng, **"The new intrusion prevention and detection approaches for clustering-based sensor networks"**, in 2005 IEEE Wireless Communications and Networking Conference, WCNC 2005: Broadband Wirelss for the Masses - Ready for Take-off, Mar 13-17 2005.

[4]. A. Agah, S. Das, K. Basu, and M. Asadi, "**Intrusion detection in sensor networks: A non-cooperative game approach**", in 3rd IEEE International Symposium on Network Computing and Applications, (NCA 2004), Boston, MA, August 2004, pp. 343346.

[5]. A. da Silva, M. Martins, B. Rocha, A. Loureiro, L. Ruiz, and H. Wong, "**Decentralized intrusion detection in wireless sensor networks"**, Proceedings of the 1st ACM international workshop on Quality of service & security in wireless and mobile networks- 2005.

[6]. OTran Hoang Hai, Faraz Khan, and Eui-Nam Huh, "**Hybrid Intrusion Detection System for Wireless Sensor Network**", ICCSA 2007, LNCS 4706, Part II, pp. 383–396, 2007. Springer-Verlag Berlin Heidelberg 2007

[7] C. Karlof and D. Wagner, "**Secure routing in wireless sensor networks: Attacks and countermeasures**", In Proceedings of the 1st IEEE International Workshop on Sensor Network Protocols and Applications (Anchorage, AK, May 11, 2003).

[8] National Institute of Standards and Technology, "Wireless ad hoc sensor networks", web: http://w3.antd.nist.gov/wahn_ssn.shtml, retrieved 12th January,2008.

[9]. Sumit Gupta  "**Automatic detection of DOS routing attach in Wireless sensor network**" MS thesis,  Faculty of the Department of Computer Science University of Houston , December 2006

[10 ] Rodrigo Roman, Jianying Zhou , Javier Lopez,  "**Applying Intrusion Detection Systems to wireless sensor networks** ",Consumer Communications and Networking Conference, 2006. CCNC 2006. 3rd IEEE, 8-10 Jan. 2006 Volume: 1,  On page(s): 640- 644 ISBN: 1-4244-0085-6

[11]. P.Bruth and C. Ko, " **Challenges in  Intrusion detection for wireless ad hoc networks**" in Application and the Internet Workshop =s, 2003 proceedings,2003 Symposium on, PP.368373,2003.

[12] R. Chadha, G. Lapiotis, S. Wright, "**Policy-Based Networking**", IEEE Network special issue, March/April 2002, Vol. 16 No. 2, guest editors.

[13] Linnyer Beatrys Ruiz Jose Marcos Nogueira and Antonio A. F. Loureiro. **MANNA: A Management Architecture for Wireless Sensor Networks**   IEEE Communications Magazine, 2003.2b: http://w3.antd.nist.gov/wahn_ssn.shtml, retrieved 1[68]. Zhou Ying, Xiao Debao, "**Mobile agent based Policy management for wireless sensor network**",  ISBN: 0-7803-9335-X/05, 2005 IEEE.
2th

[14] W. Chen, N. Jain and S. Singh, "**ANMP: Ad hoc Network Management protocol**", IEEE Journal on Selected Areas in Communications17(8) (August 1999) 1506-1531.

[15] Piya Techateerawat, Andrew Jennings, "**Energy Efficiency of Intrusion Detection Systems in Wireless Sensor Networks**", Proceedings of the 2006 IEEE/WIC/ACM International Conference on Web Intelligence and Intelligent Agent Technology (WI-IAT 2006 Workshops)(WI-IATW'06) 0-7695-2749-3/06

116